\documentclass[pra,groupedaddress,twocolumn,amsmath,nofootinbib]{revtex4}

\usepackage{graphics}
\usepackage{epsfig}

\usepackage{dcolumn}
\usepackage{bm}

\def\jpb{{J. Phys. B: At. Mol. Opt. Phys.}~}
\def\pra{{ Phys. Rev. A}~}

\def\prl{{ Phys. Rev. Lett.}~}

\def\jetp{{ Zh. \'{E}xp. Teor. Fiz.}~}

\newcommand{\vecr}{\mathbf{r}}

\newcommand{\vecp}{\mathbf{p}}

\newcommand{\vecA}{\mathbf{A}}

\newcommand{\vecE}{\mathbf{E}}

\newcommand{\vecv}{\mathbf{v}}

\newcommand{\beq}{\begin{equation}}
\newcommand{\eeq}{\end{equation}}

\newcommand{\ii}{\mathrm{i}}
\def\Vec#1{\left(\!\!\begin{array}{c}#1\end{array}\!\!\right)}

\usepackage{color}
\def\us#1{{\color{blue}\it US: #1}}

\begin{document}


\title{Boosting terahertz-radiation power with two-color circularly polarized\\ mid-infrared laser pulses}

\author{V.\,A.\,Tulsky}
\affiliation{Institute of Physics, University of Rostock, 18051 Rostock, Germany}
\affiliation{Max Planck Institute for the Physics of Complex Systems, 01187, Dresden, Germany}
\author{M.\,Baghery}
\affiliation{Max Planck Institute for the Physics of Complex Systems, 01187, Dresden, Germany}
\author{U.\,Saalmann}
 \affiliation{Max Planck Institute for the Physics of Complex Systems, 01187, Dresden, Germany}
\author{S.\,V.\,Popruzhenko}
 \email{sergey.popruzhenko@gmail.com}
 \affiliation{Max Planck Institute for the Physics of Complex Systems, 01187, Dresden, Germany}
 \affiliation{National Research Nuclear University MEPhI, 115409, Moscow, Russia}

\date{\today}

\begin{abstract}\noindent
A way to considerably enhance terahertz radiation, emitted in the interaction of intense mid-infrared laser pulses with atomic gases, in both the total energy and the electric-field amplitude is suggested.
The scheme is based on the application of a two-color field consisting of a strong circularly polarized mid-infrared pulse with wavelengths of $1.6\div 4\,\mu{\rm m}$ and its linearly or circularly polarized second harmonic of lower intensity.
By combining the strong-field approximation for the ionization of a single atom with particle-in-cell simulations of the collective dynamics of the generated plasma it is shown that the application of such two-color circularly polarized laser pulses may lead to an order-of-magnitude increase in the energy emitted in the terahertz frequency domain as well as in a considerable enhancement in the maximal electric field of the terahertz pulse.
Our results support recently reported experimental and numerical findings.
\end{abstract}

\pacs{\us{PACS are not supported any longer by Phys.\,Rev.\ {\rm\tt[https://journals.aps.org/PACS]}, one has to give subject headings {\rm\tt[https://physh.aps.org]} during submission}}
\maketitle

\section{Introduction}\noindent
During the last decade, a new research area of the generation and control of intense electromagnetic pulses in the terahertz (THz) frequency domain has emerged within the physics of laser-matter interaction.
The fast development of this field is stimulated by a considerable potential of THz radiation for fundamental research and applications in molecular and solid-state physics, spectroscopy of materials, noninvasive diagnostics for medical and security purposes and more \cite{applphys06,jew-prl10,hirori-nc11,watanabe-oe11,nphot11,fleischer11,ewers-prb12,katayama-prl12,optexp12,tanaka-np13,sherman-np15}.
For these applications, coherent pulses in a broad frequency range of roughly $0.1\div 100$\,THz with the electric field strength of the order of 1\,MV/cm and higher are of demand.
Several generation methods have been proposed to reach this domain of parameters.
Synchrotrons and free-electron lasers are used as efficient THz sources already for several decades (see e.g. \cite{gavrilov,abo-bakr,carr,williams} and references therein).
Their application is, however, limited by the large extents and the high costs of these devices. 
The advent of powerful table-top lasers with a high repetition rate paved the way for the development of cheap and compact laser-based THz sources.
Such sources can employ the optical rectification effect as well as other nonlinear generation mechanisms in crystals and liquids \cite{yeh-apl07,kitada-apl09,jiang-ol11,hirori-apl11,fortov,shalaby,jin-apl17}.
Besides, relativistic laser plasmas at solid-state densities were also shown to efficiently emit in the THz frequency domain.
The latter method, although restricted by a low repetition rate of multi-TW lasers, has demonstrated the presently highest peak-power of THz radiation \cite{gopal-njp12,gopal-prl13}.

A promising alternative to these methods is based on nonlinear-wave conversion in gases ionized by intense laser radiation.
These schemes involve such physical mechanisms as four-wave mixing \cite{bartel-ol05}, carrier-envelope phase effects in ultrashort pulses \cite{vved-prl09}, and excitation of asymmetric photo-electron currents by multi-color fields \cite{kim-oe07,vved-prl18}, including the generation of THz waves by two-color laser filaments \cite{wang-apl09,chin-oe12,berge-prl13}.
Apart from the high repetition rate provided by femtosecond laser systems and the absence of a damage threshold, the gaseous scheme benefits from its simplicity in implementation: ambient air can serve as a generating medium.
By now, pulses with an electric field strength of the order of 10\,MV/cm \cite{kim-apl14} have been obtained providing the presently strongest THz sources operating in the high-repetition-rate regime.
Even higher efficiencies of THz conversion leading to electric field amplitudes up to 1\,GV/cm have been recently predicted for the case when mid-infrared two-color laser pulses with a wavelength of 3.9\,$\mu$m are being used \cite{fedorov-pra18} instead of the conventional 800\,nm radiation of Ti:Sapphire lasers.

The physics underlying the multi-color and the short-pulse schemes of THz generation in gases is well understood on the level of a single-atom ionization event.
A strong laser field with an intensity of the order of $10^{14}$\,W/cm$^2$ quickly ionizes atoms or molecules creating a plasma channel.
The initial velocity and coordinate distribution $f_0(\vecp,\vecr)$ of this plasma is determined by the probability of ionization into a state with momentum $\vecp$ of an atom placed at position $\vecr$.
In a monochromatic electromagnetic field, photo-electron momentum distributions possess inversion symmetry, so that $f_0(\vecp,\vecr)=f_0(-\vecp,\vecr)$ resulting from the respective symmetry of the field, $\vecE(t+T/2)=-\vecE(t)$, with $T=2\pi/\omega$ being the optical period.
For non-monochromatic fields this does not hold any longer, even if they remain periodic, making the distribution asymmetric in the momentum space.
This asymmetry results in a non-zero net photo-electron current ${\bf j}(\vecr,t)$ which evolves in time and emits radiation.
The macroscopic dynamics of the ionization-prepared plasma is usually described within approximate schemes employing different model expressions for the photo-electron current.
Examples can be found in Refs.\,\cite{kim-pp09,vved-prl14}.

The two-color scheme employing an infrared fundamental pulse and its second or half harmonic of lower intensity was widely used in THz experiments.
The simplest frequency ratio 1:2 is experimentally advantageous, because the generation of the second or half harmonic of considerable intensity on the level of a few percent of the fundamental is straightforward by employing either a nonlinear crystal \cite{kim-oe07} or a parametric amplifier \cite{vved-prl14}.
Application of other frequency ratios such as 2:3 or 3:4, etc.\ can also lead to the excitation of strong photoelectron currents, at least in the case when the relative amplitudes of the two fields are comparable \cite{vved-prl16}.
As far as the power of THz radiation is concerned, such technically more demanding schemes have not yet been shown to offer any significant advantage compared to the scheme employing the 1:2 ratio. 
Thus, in the following we focus of on this simplest and most common case.

Up to now, most of the studies of this scheme have been restricted to linear polarization (LP) of laser pulses \cite{kim-oe07,kotelnikov-jetp11,you-prl12,kim-pra13,vved-prl14,poprz-pra15}.
Much less attention has been paid to the case of elliptically (EP) and circularly polarized (CP) fields, though EP and CP pulses were used to control polarization properties of THz radiation \cite{dai-prl09,wen-prl09}.
Recently, it was demonstrated \cite{meng-apl16} that application of CP pulses can lead to a considerable enhancement in the energy emitted in the THz domain, compared to the case of LP with all other parameters fixed. 
Applying two CP pulses consisting of an 800\,nm fundamental field of intensity $(1\div 5)\cdot 10^{14}$W/cm$^2$ and its second harmonic (SH) with an intensity on the level of 5\% of the fundamental, a five-fold increase of the energy emitted in the THz domain, compared to the case when the same pulses were LP, was documented.
This observation opens a way to further increase the IR-to-THz conversion efficiency.
From the other side, recent experiments performed with mid-infrared pump pulses of wavelengths $1.8\div 3.9\,\mu$m \cite{clerici-prl13,baltushka-cleo18} have also demonstrated a considerable enhancement in the power of the THz radiation, compared to the case of an 800\,nm pump laser field.

In this paper, we suggest to combine the advantages of longer wavelengths \cite{fedorov-pra18,berge-prl13,clerici-prl13,baltushka-cleo18} and of circular polarization \cite{meng-apl16} which have been separately studied in experiment and shown to increase the efficiency of conversion of laser radiation into THz waves.
To this end we examine theoretically the THz generation in a two-color field consisting of a fundamental CP and a second-harmonic field that is either CP or LP at different laser wavelengths. 
Our analysis covers both the single-atom ionization dynamics and the collective motion of the created electron plasma.
The results confirm that the fundamental-field wavelength is a crucial parameter which determines the THz radiation power, the electric field amplitude and, partially, the duration of the THz pulse.
Our calculations predict that the application of mid-infrared CP pump pulses with wavelengths $1.6\div 4\,\mu{\rm m}$ may lead to a considerable increase in the emitted THz energy and a several-fold enhancement in the maximal strength of the generated electric field.
Atomic units are used unless noted otherwise.

\section{Single-atom response}\noindent 
We start by analyzing the single-atom response which forms the initial distribution of the electron plasma.
To this end we calculate photo-electron momentum distributions in the dipole approximation in a two-color laser field with the vector potential 
\begin{widetext}
\beq
\vecA(t)=\vecA_{\omega}(t)+\vecA_{2\omega}(t)=p_F\,g(t)\bigg\{\Vec{\sin\omega t\\ \cos\omega t\\ 0}+
\frac{b}{2\sqrt{1+\xi^2}}\Vec{\sin(2\omega t{+}\alpha)\\ \xi\cos(2\omega t{+}\alpha)\\ 0}\bigg\}~,
\qquad p_F=\frac{E_0}{\omega}.
\label{At}
\eeq
\end{widetext}
Here $p_F$ is the characteristic momentum in the monochromatic electromagnetic field of amplitude $E_0$ and frequency $\omega$, $b$ is the SH fraction, $\alpha$ is the relative phase shift and $\xi$ is the ellipticity of the SH field, for which we consider $\xi{=}{\pm}1$ (CP) and $\xi{=}0$ (LP), respectively.
The envelope function $g(x)$ describes pulses with duration of $N$ fundamental periods
\beq
g(t)=1-\sin^4\Big(\frac{\omega t}{2N}\Big),
\label{f}
\eeq
acting in the time interval $\omega t\in [-N\pi,N\pi]$.
The advantage of the scheme with a CP fundamental pulse was explained before \cite{meng-apl16} using a tunneling-ionization model. 
In LP fields the characteristic photo-electron momentum is of the order of $\sqrt{F}\,p_F$ while in CP it is close to $p_F$.
Here $F=E_0/E_{\rm ch}$ is the reduced laser field, which is the original amplitude $E_{0}$ scaled by the characteristic field of the atomic level $E_{\rm ch}=(2I_{\rm p})^{3/2}$ with $I_{\rm p}$ the ionization potential.
Ionization proceeds efficiently already in laser fields with $F\ll 1$ \cite{popov-ufn04,poprz-jpb14}.
For example, ionization of xenon and argon with $I_{\rm p}{=}12.1$\,eV and 15.8\,eV, respectively, used in \cite{meng-apl16}, requires intensities of ${\cal I}\sim (1\div 5)\cdot 10^{14}$\,W/cm$^2$
resulting in corresponding reduced laser fields of $F\simeq 0.03\div 0.08$.
As a consequence, typical photo-electron momenta in CP fields are considerably higher than in LP, where they are suppressed by a factor $\sqrt{F}$ and the absolute difference grows with the wavelength \cite{popov-ufn04}.
In a monochromatic field photo-electron angular distributions possess inverse symmetry, so that the net momentum is equal to zero both for LP and CP.
When a weak second-harmonic pulse destroys the symmetry of distributions, the resulting net momentum is of the order of $b\,p_F$ for LP \cite{kim-pp09,kotelnikov-jetp11} and $p_F$ for CP pulses, leading to the generation of a stronger electric current in the latter case.
This idea was qualitatively explored and experimentally verified \cite{meng-apl16} without analyzing the subsequent collective electron dynamics.
The same observation suggests the main idea of the present paper, namely using CP mid-infrared pulses with $p_F$ several times higher at the same intensity in order to achieve stronger currents and therefore a higher power of THz radiation.
As the net drift-momentum of photo-electrons scales linearly with the wavelength, $\langle p\rangle\approx p_F\sim\lambda$, the initial dipole moment of the plasma and the amplitude of the generated electric field are expected to follow the same scaling \cite{clerici-prl13}.
\begin{figure}[t]
\includegraphics[width=8cm]{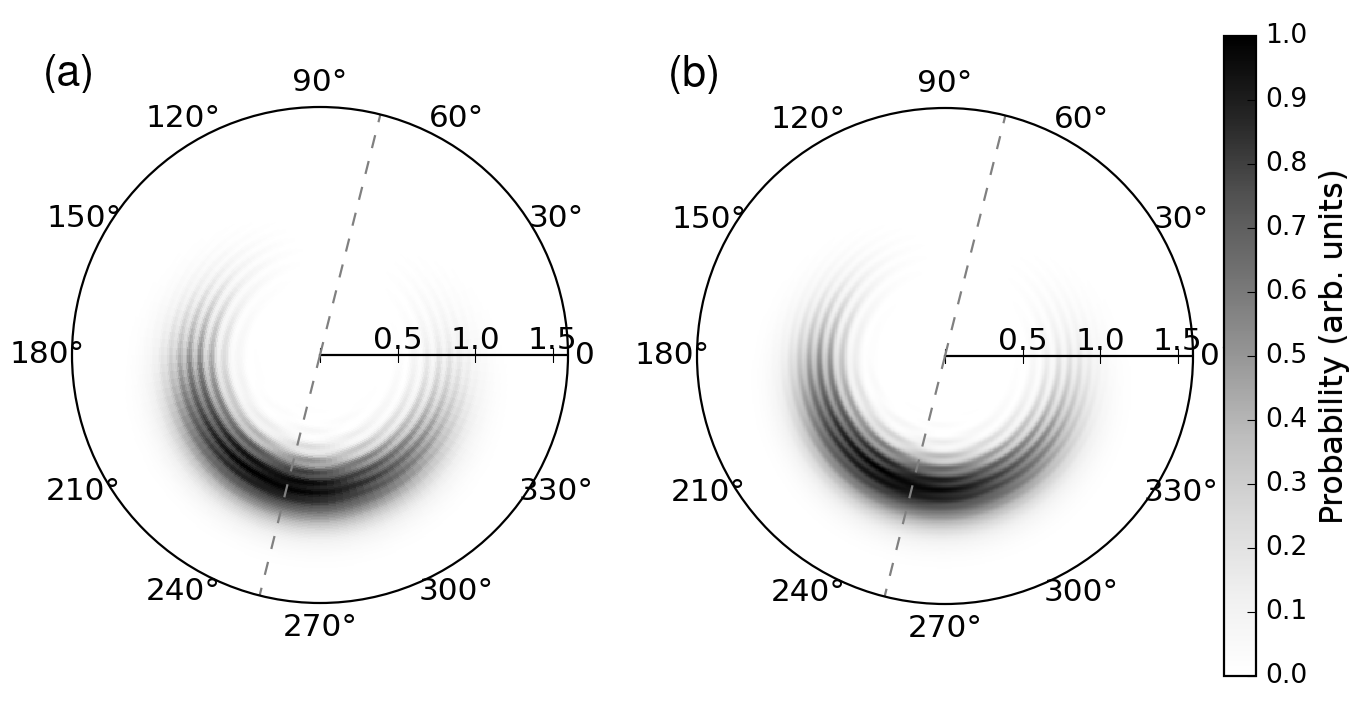}
\caption{\label{fig:tdse-tdse} Momentum distributions of photo-electrons in the polarization plane for ionization of hydrogen in a 3-cycle 800nm+400nm laser pulse of the total intensity ${\cal I}=10^{14}$W/cm$^2$ shared between the fundamental field and the SH as 0.95~:~0.05 (b=0.22).
Panel (a): distribution calculated using the code QPROP \cite{qprop}, panel (b): the same distribution found using the code \cite{patch}.
Dashed lines indicate the direction of the net photo-electron momentum $\langle\vecp\rangle$. Horizontal grid line shows photo-electron momenta in atomic units.}
\label{fig:2d-dist-tdse}
\end{figure}%
\begin{figure}[t]
\includegraphics[width=8cm]{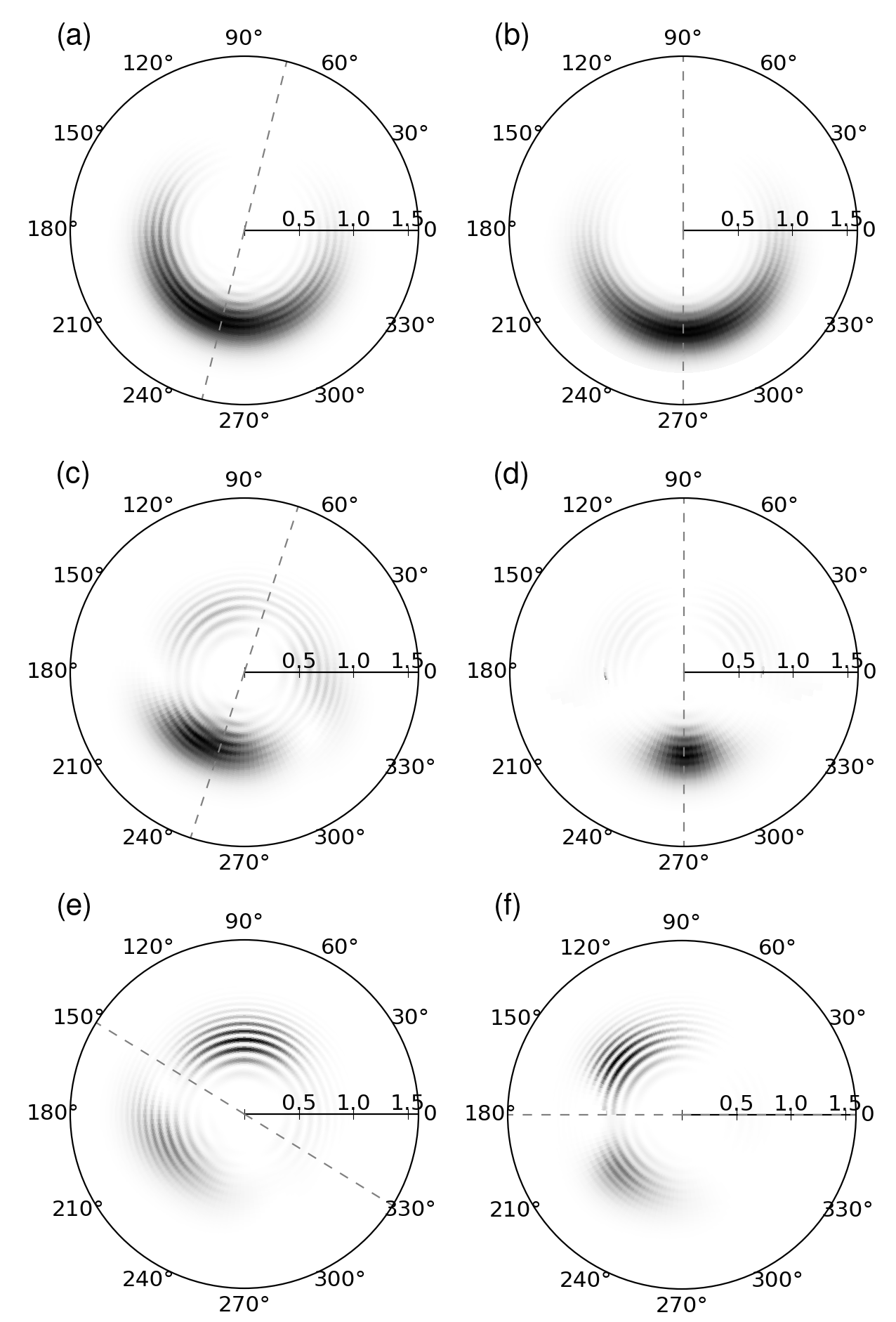}
\caption{\label{fig:tdse-sfa} Photo-electron momentum distributions in the polarization plane calculated for the ionization of hydrogen in a 3-cycle 800\,nm\,+\,400\,nm laser pulse using either the TDSE (a,c,e) or the SFA (b,d,f).  
In all cases the fundamental field is CP with intensity $0.95\cdot 10^{14}$W/cm$^2$ and the SH field is of intensity $5\cdot 10^{12}$W/cm$^2$ with CP, $b=0.22$ (a,b), LP, $b=0.32$ with $\alpha=0$ (c,d) and LP with $\alpha=-\pi/2$ (e,f).
Axes along the direction of the net momentum $\langle\vecp\rangle$ are shown by dashed lines. 
Horizontal grid line shows photo-electron momenta in atomic units. 
The grayscale is the same as on Fig.\,\ref{fig:2d-dist-tdse}.}
\label{fig:2d-dist-tdse-sfa}
\end{figure}%

\begin{figure*}
\includegraphics[width=18cm]{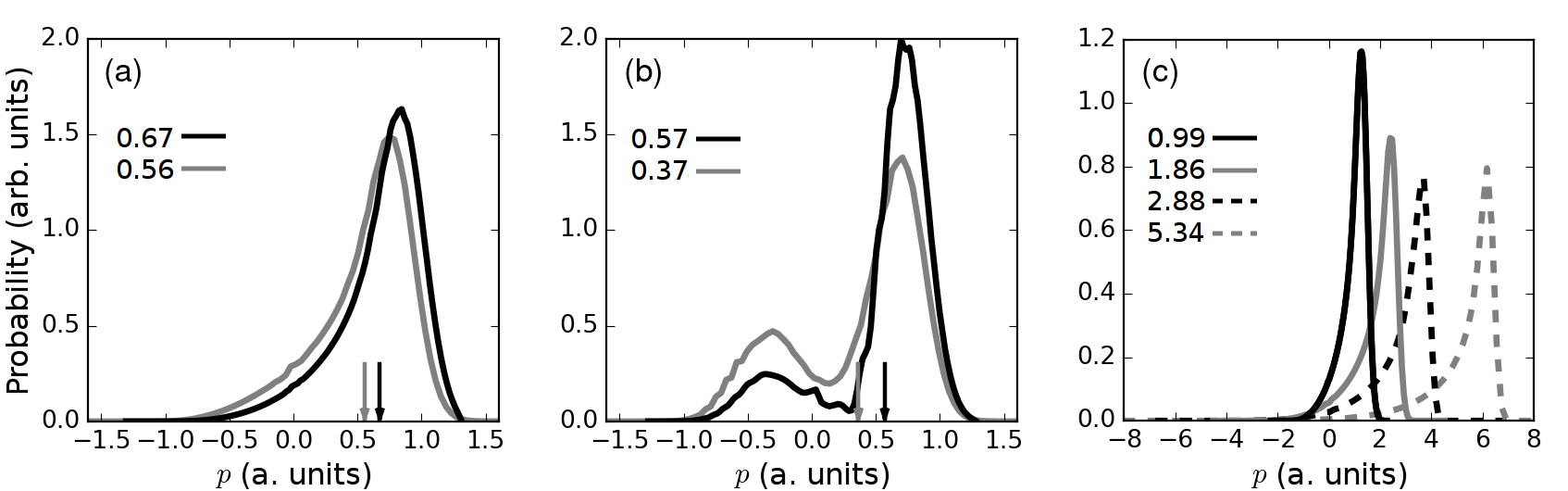}
\caption{\label{fig:integrated} Distributions in the momentum component parallel to the net momentum $\langle\vecp\rangle$ (whose direction is shown by dashed lines on Fig.\,\ref{fig:2d-dist-tdse-sfa}) integrated along the lateral momentum component for the case of Fig.\,\ref{fig:2d-dist-tdse-sfa}(a,b) (panel (a)) and Fig.\,\ref{fig:2d-dist-tdse-sfa}(c,d) (panel (b)). The SFA results are shown by black lines, the TDSE -- by grey lines. The corresponding values of the net photo-electron momentum are shown by arrows on panels (a) and (b) and by numbers on each panel. For this case $p_F=0.64$\,a.u. Panel (c) shows the SFA integrated distributions for ionization of argon with $I_{\rm p}=15.8$\,eV at intensity $3\cdot 10^{14}$W/cm$^2$ and fundamental wavelengths 800\,nm (black solid line), 1.6\,$\mu$m (gray solid line), and 2.4\,$\mu$m (black dashed line) and 4.0\,$\mu$m (grey dashed line).}
\label{fig:integ-dist}
\end{figure*}%

In obtaining theoretical photo-electron momentum distributions in mid-infrared laser fields with $\lambda=2\div4\mu{\rm m}$, we can only rely on approximate analytic approaches since exact solutions of the time-dependent Schr\"odinger equation (TDSE) in CP fields are almost out of reach at such wavelengths.
The Keldysh theory or the strong-field approximation (SFA) \cite{keldysh,faisal,reiss,popov-ufn04,poprz-jpb14} are routinely used for approximate calculations of photo-electron spectra.
In order to benchmark our SFA calculations we first consider ionization of a hydrogen atom by a 3-cycle 800\,nm laser pulse superimposed by its SH, as defined in Eq.\,\eqref{At}.
For these parameters TDSE calculations take a reasonable time at intensities ${\cal I}\simeq 10^{14}$W/cm$^2$.
To enhance the reliability of numerical results we used two independently developed TDSE solvers \cite{qprop,patch}.
Convergency in the calculation of momentum distributions was achieved by accounting angular momenta $\ell$ up to $\ell_{\rm max}=50$ in the expansion of the wave function in spherical harmonics.
Distributions shown on Fig.\,\ref{fig:2d-dist-tdse} clearly demonstrate a good quantitative agreement between results obtained with both different Schr\"{o}dinger solvers.

As a next step we compare photo-electron distributions obtained with QPROP \cite{qprop} for several two-color laser pulses with those calculated within the SFA.
The latter was used in its simplest version when no Coulomb effects are accounted for.
In this case the SFA ionization amplitude is given by \cite{poprz-jpb14}
\begin{subequations}\label{eq:sfa}
\beq
M(\vecp)=\ii\int\limits_{-\infty}^{+\infty} dt\int d^3r\,\Psi_{\vecp}^*(\vecr,t)\,\vecE(t)\cdot\vecr\,\Psi_0(\vecr)e^{\ii I_{\rm p}t}.
\label{Mp}
\eeq
Here $\Psi_0$ is the bound state wave function and the photo-electron continuum is approximated by plane Volkov waves 
\beq
\Psi_{\vecp}(\vecr,t)=\frac{1}{(2\pi)^3}\exp\Big(\ii\vecv_{\vecp}(t)\cdot\vecr-\frac{\ii}{2}\int\limits^t\vecv_{\vecp}^2(t')dt'\Big),
\label{VPsi}
\eeq
where
\beq
\vecv_{\vecp}(t)=\vecp+\vecA(t)
\label{v}
\eeq
\end{subequations}
is the time-dependent photo-electron kinematic momentum with the asymptotic value $\vecp$ at a detector.
The spatial integrals in the matrix element of (\ref{Mp}) are calculated analytically, and the time integral by the steepest-descent method with complex-valued saddle points $t_{\rm s}$ are found numerically from the equation 
\beq
\big[\vecp+\vecA(t_{\rm s})\big]{}^2+2I_{\rm p}=0.
\label{spe}
\eeq
The photo-electron momentum distribution is then
\beq
dw(\vecp)=\big|\sum_s M(\vecp,t_{\rm s})\big|^2\,d^3p
\label{dw}
\eeq
Some results are summarized in Figs.\,\ref{fig:2d-dist-tdse-sfa} and \ref{fig:integ-dist}.
Figure~\ref{fig:2d-dist-tdse-sfa} shows photo-electron momentum distributions in the polarization plane calculated from TDSE using the QPROP \cite{qprop} and from the SFA along Eqs.\,\eqref{eq:sfa}--\eqref{dw} for different configurations of the SH field.
In the case of the circularly polarized SH (panels (a) and (b)) a quantitative agreement is apparent.
This is supported by the plots of Fig.\,\ref{fig:integ-dist}(a), where the projected distributions along the direction of the net momentum $\langle\vecp\rangle$ (i.\,e.\ integrated over the momentum perpendicular to this direction) are shown. 
Note, that the distributions are not exactly symmetric with respect to the axes shown by dashed lines, as is particularly evident for the cases of Fig.\,\ref{fig:2d-dist-tdse-sfa}(c-d).
This weak asymmetry results from the short pulse duration and vanishes when longer pulses are applied as is done in Section III.
Here the deviation of the SFA result from the exact one does not exceed 20\%.
The overall rotation of the TDSE distributions with respect to those calculated using the SFA is a well-known Coulomb effect \cite{gor-prl04,smirnova-np15}, which makes neither a significant impact on the absolute value of the average momentum nor on the width of the distribution.
This Coulomb rotation will, however, affect the angular distribution and polarization of the emitted THz radiation.

For the case when the SH is linearly polarized with a relative phase shift of $\alpha\,{=}\,0$ (panels (c),\,(d) in Fig.\,\ref{fig:2d-dist-tdse-sfa}) and $\alpha\,{=}\,{-}\pi/2$ (panels (e),\,(f)) the agreement is considerably poorer and stays on the qualitative level only.
The important message of these results is that for LP the SFA predicts the distributions to be sharper than they appear in the exact calculation.
As a consequence, as far as the net photo-electron momentum is concerned, the scheme with LP appears less advantageous, which becomes even clearer from the numbers given in Fig.\,\ref{fig:integ-dist}.
Moreover, the shape of the momentum distribution in the CP+LP combination depends strongly on the relative phase $\alpha$ in (\ref{At}), as it clearly seen from the distributions of Fig.\,\ref{fig:2d-dist-tdse}(c--f), making the output sensitive to fluctuations of this relative phase.
Therefore the results shown on Figs.\,\ref{fig:2d-dist-tdse-sfa} and \ref{fig:integ-dist} suggest that the CP+CP scheme \cite{meng-apl16} provides the most promising initial condition for the subsequent plasma dynamics by forming a strong photo-electron current insensitive to a relative phase $\alpha$.
In the following we focus on the case when both fields are circularly polarized, while the CP+LP scheme will be analyzed in details elsewhere.

The accuracy of the SFA is known to increase in the regime of tunneling when the Keldysh parameter \cite{keldysh}
\beq
\gamma=\frac{\sqrt{2I_{\rm p}}\,\omega}{E_0}
\label{gamma}
\eeq
becomes smaller than unity.
For the case when argon is used as a target gas and the laser intensity at the fundamental frequency is taken to be close to $3\cdot 10^{14}$W/cm$^2$ we obtain $\gamma=0.93/k$ with $k=1,2,3$ and 5 for $\lambda=800$\,nm, 1.6\,$\mu$m, 2.4\,$\mu$m and 4.0\,$\mu$m, respectively.
Thus, in application to mid-infrared fields within the CP+CP scheme we are quite confident that our SFA results will reproduce exact ones even better than they do for the parameters of Fig.\,\ref{fig:integ-dist}(a).
Note that in the case of counter-rotating CP pulses, i.\,e.\ $\xi{=}{-}1$ in Eq.\,(\ref{At}), the distribution possess a three-fold symmetry \cite{milosevic}, the net momentum vanishes, and no THz emission is expected, in agreement with experiment \cite{meng-apl16}.

\section{Collective dynamics of the electron plasma and radiation}\noindent
In order to examine radiation spectra emitted by this preformed plasma we adopt the following model.
We assume that the bichromatic pump pulse sets up the electron plasma instantaneously.
Indeed, a frequency of $\nu=10$\,THz corresponds to a time scale of $1/\nu\simeq 100$\,fs, that is four times longer than the pulse duration in the experiment \cite{meng-apl16}. 
Thus, the frequency domain we consider here ($0.1\div 10$\,THz) involves timescales exceeding the pulse duration by roughly one order in magnitude or even more.
This means that THz radiation is mostly emitted after the interaction of the gas with the laser pulses and the spectrum is determined by free plasma oscillations induced in the ionized gas.
Note that for parameters different from that we consider (including longer laser pulses) this may not be the case. 

A second simplification we apply consists in restricting the problem to an 1D geometry.
To this end, we fix the coordinate $z$ along the propagation direction of the laser pulse and integrate the initial distribution in the polarization plane in momenta perpendicular to the direction of $\langle\vecp\rangle$ shown in Figs.\,\ref{fig:tdse-tdse} and \ref{fig:tdse-sfa}.
The resulting 1D momentum distributions are shown in Fig.\,\ref{fig:integrated}(c) and briefly discussed above.
In this approximation the collective electron dynamics evolves in a 2D phase-space $(x,p)$, and the plasma layers, corresponding to different $z$, are considered independently.
The latter approximation is justified by the relatively small transverse spread of photo-electron-momentum distributions, $\Delta p_z\simeq\sqrt{I_{\rm p}F}\ll \sqrt{I_{\rm p}}$ and $\Delta p_z\ll p_F$, so that the plasma can be considered cold in this direction.
Contributions from different $z$ may add up coherently in the radiation spectrum if the phase difference between the emitted fields remains numerically small across the efficiently emitting area whose size is determined by an experimental setup.

In the experiment by Meng et al.\ \cite{meng-apl16} a gas-filled tube with a diameter of 100\,$\mu$m has been used.
This size is comparable to the wavelength of the emitted THz radiation and much smaller than the Rayleigh length of a weakly focused laser beam.
In this situation, when the gas medium is restricted in all three directions and emits approximately as a dipole, effects of propagation are of secondary importance.
A clear quadratic dependence of the THz power on the gas pressure observed in \cite{meng-apl16} below 700\,mbar supports the conclusion that the ionization-preformed plasma coherently radiates when the interaction volume is restricted to a small region.
In this paper, we remain focused on such setup.
In an alternative case when laser pulses interact with a gas in a chamber or with ambient air, filaments of the length up to several cm and even longer can be formed \cite{fedorov-pra18,chin-oe12,berge-prl13,zheltikov-sr15}.
For a correct description of this experimental realization, effects of propagation must be taken into account both for the THz signal and for the two-color laser pulse itself.
\begin{figure}[b]
\includegraphics[width=6cm]{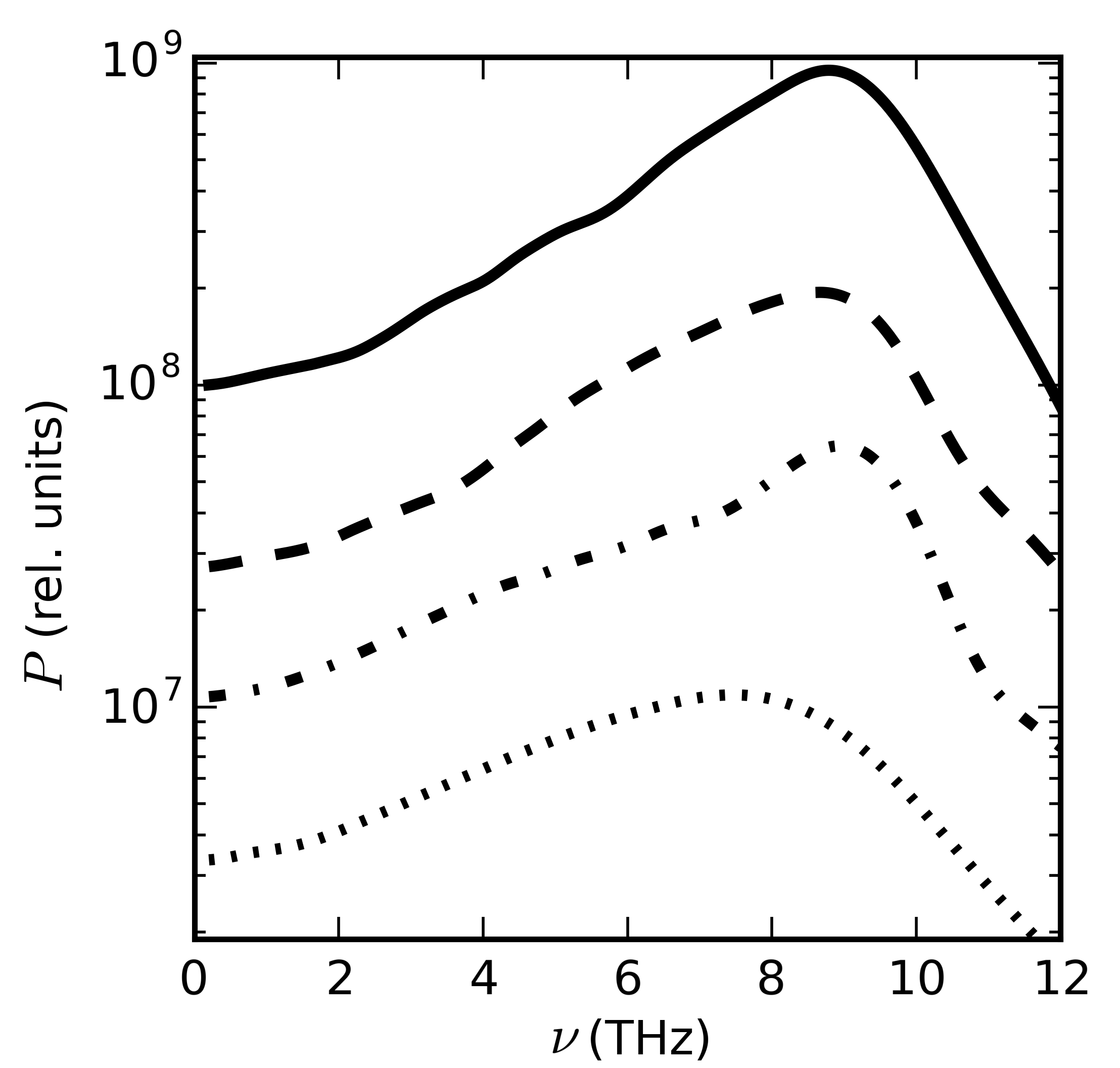}
\caption{Spectral density of the radiated power calculated according to (\ref{P}) for an argon gas at a backing pressure of 200\,mbar and laser intensity of $3\cdot10^{14}$W/cm$^2$ for the fundamental wavelengths of $\lambda\,{=}\,800$\,nm (dotted line), 1.6\,$\mu$m (dash-dotted line), 2.4\,$\mu$m (dashed line) and 4\,$\mu$m (solid line).}
\label{fig:spectral} 
\end{figure}%


Under these assumptions the plasma dynamics after the interaction with the laser can be modeled by means of a 1D electrostatic Poisson-Vlasov equation for the distribution function $f_{z}(p,x,t)$ in a plasma slab at fixed $z$
\begin{subequations}\label{eq:vlasov-poisson}
\beq
\bigg\{\frac{\partial}{\partial t}+v_x\frac{\partial}{\partial x}+E_x(x,t)\frac{\partial}{\partial p}\bigg\}f_{z}(p,x,t)=0
\label{vlasov}
\eeq
\beq
\frac{\partial}{\partial x}E_x(x,t)=4\pi\!\int\! f_{z}(p,x,t)dp-4\pi\,n_{{\rm i},z}(x)
\label{poisson}
\eeq
\end{subequations}
Here $n_{{\rm i},z}(x)$ is the charge density of the immobile ions which is equal in absolute value to the one of the electrons at the initial time instant
\beq
n_{{\rm i},z}(x)=\int\! f_{z}(p,x,t{=}0)\,dp.
\label{rho}
\eeq
Note that the ion background in Eq.\,(\ref{poisson}) is responsible for the net acceleration of the electron cloud and therefore for the dipole radiation emitted when the laser pulse is off.
The initial density profile (\ref{rho}) was calculated assuming tunneling ionization of Ar atoms by the laser field with a Gaussian distribution of intensity 
\beq
{\cal I}_{z}(x)=\frac{{\cal I}_0}{1{+}z^2/z_R^2}\exp\Big({-}\frac{2x^2}{w_0^2(1{+}z^2/z_R^2)}\Big),
\label{I}
\eeq
where $w_{0}$ is the focal radius and $z_R=\pi w_0^2/\lambda$ is the Rayleigh length.
We would like to remind here that the parametric dependence on $z$ in Eqs.\,\eqref{eq:vlasov-poisson} is due to the  fixed ion density $n_{{\rm i},z}(x)$ and the initial electron density  $n_{{\rm e},z}(x,t{=}0)={-}\!\int\! f_{z}(p,x,t{=}0)\,dp$ only. 

The coupled Eqs.\,\eqref{vlasov} and \eqref{poisson} were solved numerically by means of a particle-in-cell (PIC) code on the segment $x\in [-5w_0,+5w_0]$ with a focal radius of $w_0=10\,\mu{\rm m}$ independently for 21 values of the longitudinal coordinate $z\in [{-}100\,\mu{\rm m},0]$ with the following extension to $z\in [0,{+}100\,\mu{\rm m}]$ using a symmetry $f_{-z}(x,p,t)=f_{+z}(x,p,t)$.
No upper and lower limits in the momentum space were imposed, and the boundary condition in $x$ was taken to be periodic.
\begin{figure}[t]
\includegraphics[width=8cm]{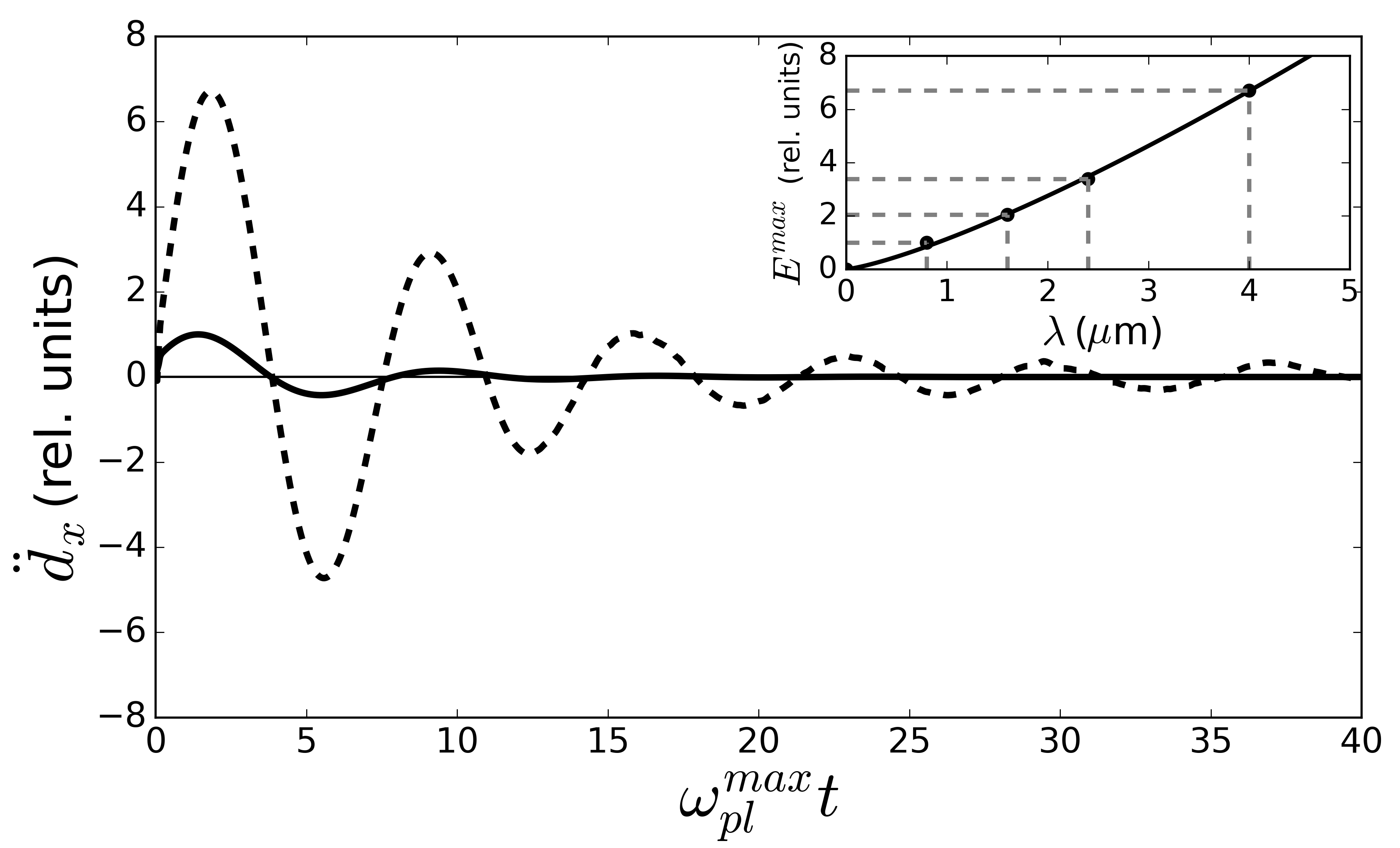}
\caption{Time dependence of the dipole acceleration (\ref{dx}) for the case of argon gas at backing pressure 200\,mbar and laser intensity $3\cdot10^{14}$W/cm$^2$ for the fundamental wavelength of 800\,nm (solid line) and 4\,$\mu$m (dashed line). The inset shows the wavelength dependence of the maximal electric field with the power-law fit from Eq.\,\eqref{E-lambda}.}
\label{fig:time}
\end{figure}%

An important point of this otherwise trivial simulation consists in the necessity to take into account electron-ion and electron-electron binary collisions whose frequency reads \cite{dau10}
\beq
\nu_{\rm ei}\simeq\nu_{\rm ee}\simeq\frac{4\pi n_{\rm e}\Lambda}{T_{\rm eff}{}^{3/2}}\equiv\nu,
\label{nu}
\eeq
with $\Lambda\simeq 5$ being the Coulomb logarithm. 
Here $\nu_{\rm ei}$ is the frequency of electron-ion collisions determined with respect to the transfer of momentum and not of energy \cite{dau10}.
It can be estimated by taking the effective temperature equal to the distribution spread in momentum space $T_{\rm eff}=(\Delta p)^2/2$, with the latter extracted from the SFA momentum distributions, cf.\ Fig.\,\ref{fig:integ-dist}(c).
For the initial plasma density on the order of $n_{\rm e}=10^{18}\div10^{19}$cm$^{-3}$, typical for THz experiments in two-color fields and $\Delta p\approx (0.3\div 1.0)$a.u. this frequency varies 
in the interval $10^{11}\div 10^{13}\,\mbox{s}^{-1}$ which agrees with the time scale of $0.1\div 10$\,ps, that is responsible for the emission of THz frequencies.
This means that a relatively fast damping of the plasma oscillations may have a significant effect on the radiation spectra.
When collisions are taken into account, the equations of motion for the macro-particles read
\beq
\ddot{x}_a(t)=-E_x(x_a,t)-\nu\,\dot{x}_a(t).
\label{newton}
\eeq
The value in Eq.\,(\ref{nu}) scales with charge $Q_a$ and mass $M_a$ of the macro-particle such that Eq.\,(\ref{newton}) does not depend on the macro-particle size.
Energy conservation has been checked in the case $\nu=0$.

At the same time, we neglect elastic and inelastic electron-atom collisions.
At photoelectron energies in the interval $\varepsilon\approx 20\div 500$eV corresponding to the momentum distributions of Fig.\,3(c), the cross sections of elastic scattering and impact ionization are close to the geometrical cross section of the atom.
As an example, the impact ionization cross section of neutral argon does not exceed the value of $\sigma_{\rm max}\approx 3\cdot 10^{-16}$cm$^2$ \cite{fletcher-jpb73}.
Taking for an estimate the concentration of neutral atoms survived after the field ionization equal to $n_{\rm a}\simeq 10^{18}$cm$^{-3}$ and the same for the concentration of electrons, we obtain that impact ionization can increase the concentration of electrons by $\sigma_{\rm max}n_{\rm_a}v_0\tau\approx 0.1$ of its initial value during the time interval $\tau\,{=}\,1$ps.
Here $v_0$ is the average electron velocity which can be extracted from the data of Fig.\,3.
The numbers taken for this estimate correspond to the parameters of the simulation described below and give an upper limit for the contribution of impact ionization to the generation of an electron plasma.
A slow variation of the electron density on a level not exceeding 10\,\% of the initial value can hardly make a qualitative impact on the plasma oscillations.

The main factors determining the plasma dynamics are the initial distribution in the $(x,p)$--space, the value of the local plasma frequency $\omega_{{\rm pl},z}(x)=\sqrt{4\pi\,n_{{\rm e},z}(x)}$ and that of the collision frequency (\ref{nu}).
These factors can be controlled via wavelength, pulse duration, and intensity, which determine the ionization degree, and thereby $n_{\rm e}$, and the initial distribution, and finally  $T_{\rm eff}$.
The plasma frequency varies from the maximal value $\omega_{\rm pl}^{\rm max}$ at the center of the focal spot to zero at the edge.
For argon at an intensity of $3\cdot 10^{14}$\,W/cm$^2$ and a pulse duration of 40\,fs (as taken in our calculations) tunneling ionization rates \cite{popov-ufn04} predict the ionization degree to be about 40\,\%, giving $\omega_{\rm pl}^{\rm max}\approx 6\cdot 10^{13}$\,s$^{-1}$, i.\,e.\ $\nu^{\rm max}\approx 10$\,THz, at 200\,mbar backing pressure. 
Varying the pulse duration one can control the ionization degree and therefore the value of $\omega_{\rm pl}^{\rm max}$.
Equivalently, this can be done by varying either the gas density in the target or the laser intensity.

Oscillations proceed with the local plasma frequency $\omega_{{\rm pl},z}(x)$ leading to a relatively broad radiation spectrum with a maximum around of $\nu^{\rm max}$ as shown in Fig.\,\ref{fig:spectral} for various driver wavelengths $\lambda$.
At higher frequencies ($\nu\,{>}\,\nu^{\max}$) the radiation spectrum drops down quickly, while for $\nu\,{<}\,\nu^{\max}$ it decreases much slower, due to the contributions coming from the shoulders of the focal spot.
Collisional absorption leads to a relatively quick decay of the oscillations (particularly for shorter wavelengths for which the initial spread $\Delta p$ and therefore the temperature $T_{\rm eff}$ are smaller, see Fig.\,\ref{fig:integ-dist}(c)) and therefore to even higher broadening of the spectrum.
\begin{figure}[b]
\includegraphics[width=8cm]{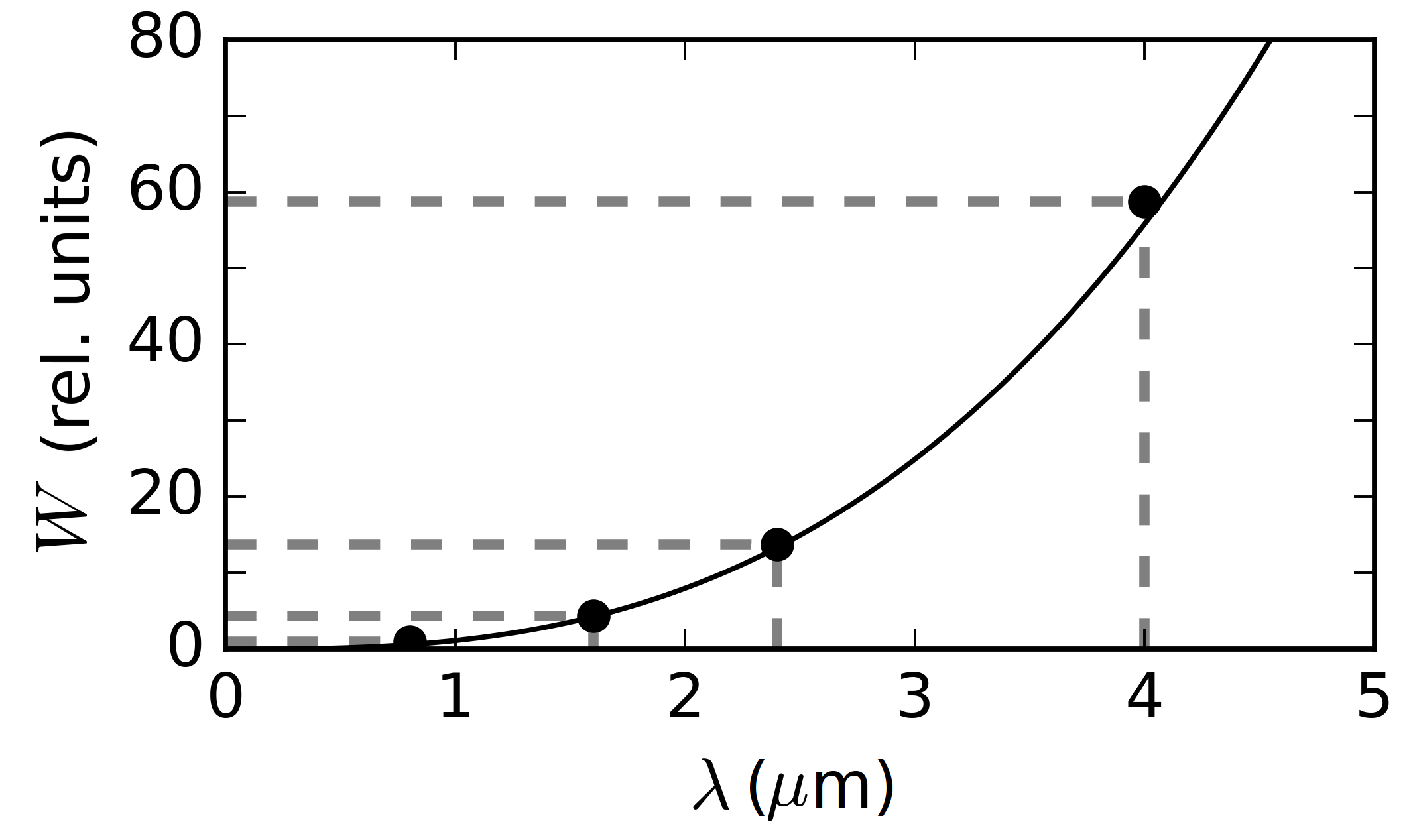}
\caption{\label{fig:power} 
Total energy emitted in the spectral range $\Delta\nu=0.1\div 10$THz as a function of the laser wavelength. The black curve shows a fit by the power law from Eq.\,(\ref{W-lambda}).}
\label{fig:energy}
\end{figure}%

Figure~\ref{fig:time} shows the time-dependent dipole acceleration of the whole plasma cloud in the interaction volume:
\beq
\ddot{d}_x(t)=-\Delta y\int\!\! dz\!\int\!\!dx\, n_{{\rm e},z}(x,t)\,\ddot{x}(t),
\label{dx}
\eeq
where $\Delta y=100\,\mu{\rm m}$ is the plasma width in the polarization plane perpendicular to the direction of plasma oscillations.
This value is the same for all wavelengths and, as we only compare relative values of the dipole moment and emitted powers, it does not influence the results.
The electric field of emitted THz waves is proportional to (\ref{dx}), so that relative values of the electric field amplitude $E^{\rm max}(\lambda)$ can be extracted from the calculations.
This dependence, shown as inset of Fig.\,\ref{fig:time}, can be nicely fitted by the power law
\beq
E^{\rm max}(\lambda)\sim\lambda^{1.28},
\label{E-lambda}
\eeq
which agrees well with the wavelength scaling of the initial electron current $|{\bf j}|\sim p_F\sim\lambda$ \cite{clerici-prl13}.

The relative spectral power of the generated radiation
\beq
P(\nu)\sim \vert\ddot d_x(\nu)\vert^2~,~~~\ddot{d}_x(\nu)=\int\limits_0^{\infty}\ddot{d}_x(t)e^{-i2\pi\nu t}dt,
\label{P}
\eeq
as shown in Fig.\,\ref{fig:spectral}, gives after integration over the frequency range $\Delta\nu=0.1\div 10$THz the total emitted energy $W$ whose dependence on the wavelength is shown in Fig.\,\ref{fig:energy}.
It fits by the power law
\beq
W(\lambda)\sim\lambda^{2.8},
\label{W-lambda}
\eeq
which can be easily understood taking into account that the total emitted energy is proportional to the square of the electric field (\ref{E-lambda}) and to the THz pulse duration which grows slowly with the wavelength due to the decrease in the efficient collision frequency (\ref{nu}).
For the case when the gas is not confined in space, but fills the whole focal volume, an additional factor of $\lambda^2$ would emerge in (\ref{W-lambda}), because of the wavelength dependence of the focal-waist radius $w_0\,{\sim}\,\lambda$, if the same optic is used for focusing at different wavelengths $\lambda$. 
In this case one would have $W(\lambda)\,{\sim}\,\lambda^{4.8}$, which agrees with the experimental power law of $\lambda^{4.6\pm0.5}$ \cite{clerici-prl13}.

\section{Conclusions}\noindent
We have studied the collective plasma dynamics and radiation initiated in a gas target by a strong two-color CP femtosecond laser pulse.
The restriction to CP in both contributing colors eliminates any dependence on the initial relative phase $\alpha$ of the two fields.
In agreement with numerical results, our qualitative estimates show that the application of mid-infrared CP pump pulses may provide an order-of-magnitude increase in the conversion efficiency of infrared laser energy into THz waves, compared to that presently achieved with 800\,nm lasers and CP \cite{meng-apl16} or with $1.8\div 3.9\,\mu$m pulses and linear polarization \cite{clerici-prl13,baltushka-cleo18}.
This enhancement results from (i) an increase in the THz pulse duration, due to the reduced frequency of binary collisions, and (ii) an increase in the electric-field amplitude, due to the higher initial value of the photo-electron current.
The latter is of exceptional importance for many practical applications of THz radiation.

Note, that several factors can limit the wavelength-dependent increase of the THz signal under realistic experimental conditions.
Firstly, the size of the focal spot grows with the wavelength if the same lens is used.
This leads to a decrease in the peak intensity and ultimately, to a fast drop in the plasma density as was observed \cite{clerici-prl13}.
Secondly, the frequency of inelastic electron-atom collisions grows with photoelectron velocity, in contrast to the electron-ion collision frequency (\ref{nu}), which decreases quickly.
As a consequence, with further increasing laser wavelength and intensity, inelastic collisions may start playing a significant role, leading to a faster stochastization of the initially coherent collective electron motion.
Finally, it is instructive to note that polarization of the THz field can also be used as an independent measure of the Coulomb rotation effect in photoelectron momentum distributions \cite{gor-prl04,smirnova-np15}.

Our conclusions are robust with respect to the assumptions made in the 1D model of the collective plasma dynamics, but restricted to the case when the interaction volume is limited by a size of about 100\,$\mu$m.
Propagation effects as well as 3D geometries may be taken into account by means of full-dimensional electromagnetic Vlasov-Maxwell solvers with microscopic currents calculated using the model presented here.

\begin{acknowledgments}\noindent
We authors are thankful to A.\ Gopal and G.\,G.\ Paulus for fruitful discussions, comments and suggestions.
S.\,V.\,P.\ who formulated the problem, performed numerical estimates and contributed into the text preparation, acknowledges financial support of the Russian Science Foundation, grant No.\,18-12-00476.
\end{acknowledgments}

\end{document}